\documentclass[journal]{IEEEtran}
\usepackage{braket}

\ifCLASSINFOpdf
\else
\fi
\usepackage{amsmath}

\begin{document}

%
\title{Analysis of Free Will and Determinism in Physics.}
%
%
%

\author{{\bf Edgar~Jos\'e~Candales~Dugarte}\\Departamento de Ciencias Aplicadas, Facultad de Ingenieria.\\Universidad de Los Andes}
\maketitle

\begin{abstract}
It is considered the study of determinism in the theories of physics. Based on fundamental postulates of physics, it is proved that the evolution of the universe is univocally determined, proving ultimately that free will does not exist. In addition, it is presented some contradictions and weaknesses of quantum mechanics, suggesting paradoxes in the theory. It is also analyzed some consequences of the postulates in justice and ethics.
\end{abstract}

\begin{IEEEkeywords}
Determinism, Free Will, Quantum Mechanics, Paradox, Physics.
\end{IEEEkeywords}

%
\IEEEpeerreviewmaketitle

\section{Introduction}
%
%
%
%

	\IEEEPARstart{I}{n} the 1820's the great French mathematician Joseph Fourier noticed that statistics on the number of births, deaths, marriages, suicides, and various crimes in the city of Paris had remarkably stable averages from year to year. The mean values in a ``normal distribution" (one that follows the bell curve or ``law of errors") of statistics took on the prestige of a social law. The Belgian astronomer and statistician Adolphe Qu\'etelet did more than anyone to claim these statistical regularities were evidence of determinism.
Individuals might think marriage was their decision, but since the number of total marriages was relatively stable from year to year, Qu\'etelet claimed the individuals were determined to marry. Qu\'etelet used Auguste Comte's term ``social physics, to describe his discovery of ``laws of human nature", prompting Comte to rename his theory sociology.
Qu\'etelet's argument for determinism in human events is quite illogical. It appears to go something like this:
Perfectly random, unpredictable individual events (like the throw of dice in games of chance) show statistical regularities that become more and more certain with more trials (the law of large numbers).
Human events show statistical regularities.
Human events are determined.
Qu\'etelet might more reasonably have concluded that individual human events are unpredictable and random. Were they determined, they might be expected to show a non-random pattern, perhaps a signature of the Determiner.

Mill's godson Bertrand Russell also had no doubt that causality and determinism were needed to do science. ``Where determinism fails, science fails", he said. Russell could not find in himself ``any specific occurrence that I could call 'will'".

Charles Sanders Peirce ``that the state of things existing at any time, together with certain immutable laws, completely determine the state of things at every other time (for a limitation to future time is indefensible). Thus, given the state of the universe in the original nebula, and given the laws of mechanics, a sufficiently powerful mind could deduce from these data the precise form of every curlicue of every letter I am now writing".

As far as we know, in the beginning due to technological advances in biology and the impartial perspectives through which scientist describe the universe indicated that humans are a kind  of animals, being classified as the rest of the organisms, based on similarities of internal structures. 
	
	A rigorous analysis, from the foundations of physics, it is possible to prove that humans are not even animals or beings with the capacity to think (or decide), in fact, no being is capable of such an ability (according to physics). If the principles upon which physics is constructed are correct, a being able to decide is an impossibility, as we are a fraction of matter subjected to specific laws, the same laws used to describe a stone or any other object apply to humans as well (since humans, as we call it, does not have a privilege state in the universe). Every configuration of the universe in a given moment is predetermined by the conditions of the universe an instant before, dictating in a precise and absolute way the evolution of the entire system, inasmuch as the physical laws are strict and well defined.

	Here, It is postulated a fundamental principle which is supposed to governed the universe. Subsequently, it is proved that such a principle leads to a unique solution for the evolution of the system described by it. Also, it is discussed the consequences of determinism in matters of justice and ethics.


 
\section{Fundamental postulate.}
	{\bf Postulate:} A physical system is ruled by a principle of least action, which states that: every system is characterized by a definite function $L(q_1,q_2,...,q_s,\dot{q}_1,\dot{q}_2,...,\dot{q}_s)$, thus, the motion of the system is such that the quantity
	\begin{equation}
		S=\int_{}^{}d^n q L(q_1,q_2,...,q_s,\dot{q}_1,\dot{q}_2,...,\dot{q}_s)
	\end{equation}
	takes the least possible value, that is to say, $\Delta S=0$. 	
\section{Uniqueness of the principle of least action}
{\bf Definition:} Let $y(x)\in M$ a class of functions and let $J[y(x)]:=M\to R$ a functional. If the increment of $J[y(x)]$, which is
		\begin{equation}
			\Delta J=J[y(x)-\delta y]-J[y(x)]
		\end{equation}
		can be represented in the form
		\begin{equation}
			\Delta J=L[y(x),\delta y]+\beta[y(x),\delta y]. ||\delta y||
		\end{equation}
		where $L[y(x),\delta y]$ is a linear functional in relation to $\delta y$ and $\beta[y(x),\delta y]\to 0$ when $||\delta y||\to 0$, then the linear increment respect to $\delta y$, that is, $L[y(x),\delta y]$, is called variation of the functional and it is represented by $\delta J$. It is said, in this case, that the functional $J[y(x)]$ is differentiable in $y(x)$.
		
		 {\bf Theorem: $\Delta J$ is determined univocally (If exists).}
		Let us consider
		\begin{equation}
			J[y(x)-\epsilon \delta y]-J[y(x)]=L[y(x),\epsilon \delta y]+\beta[y(x),\epsilon \delta y].||\epsilon \delta y||
		\end{equation}
		Since the variation $L[y(x),\delta y]$ is a linear functional, then $L[y(x),\epsilon \delta y]=\epsilon L[y(x),\delta y]$. Therefore
		\begin{equation}
			L[y(x),\delta y]=\frac{J[y(x)-\epsilon \delta y]-J[y(x)]}{\epsilon}- \epsilon \beta[y(x),\epsilon \delta y]||\delta y||
		\end{equation}
		Taking the limit when $\epsilon\to 0$, it is obtained
		\begin{equation}
			L[y(x),\delta y]=\lim_{\epsilon\to 0}\frac{J[y(x)-\epsilon \delta y]-J[y(x)]}{\epsilon}
		\end{equation}
		whose solution is univocally determined, proving that the variation is unique.

		Reductio ad absurdum: Let us suppose that $\delta J$ is not unique, but there exists two linear variations $L_1[y(x),\delta y]$ and $L_2[y(x),\delta y]$ such that $L_1[y(x),\delta y]\neq L_2[y(x),\delta y]$. Thus, we have that 
		\begin{equation}
			\Delta J=L_1[y(x),\delta y]+\beta_1[y(x),\delta y]. ||\delta y||
		\end{equation}
		and
		\begin{equation}
			\Delta J=L_2[y(x),\delta y]+\beta_2[y(x),\delta y]. ||\delta y||
		\end{equation}
		Since
		\begin{equation}
			\Delta J=J[y(x)-\delta y]-J[y(x)]
		\end{equation}
		it is obtained that
		\begin{align}
			\Delta J=L_1[y(x),\delta y]+\beta_1[y(x),\delta y]. ||\delta y||&\\=L_2[y(x),\delta y]+\beta_2[y(x),\delta y]. ||\delta y||
		\end{align}
		as $||\delta y||\to 0$ 
		\begin{equation}
			L_1[y(x)]=L_2[y(x)]
		\end{equation}

which is absurd, since we had supposed that $L_1[y(x)]\neq L_2[y(x)]$. Thus, we conclude that (if exists) two variations of the functional, they are in reality the same, that is to say, the variation is unique.
\section{Demonstration of constrained will.}
Constrained will is understood as the opposite to free will.
	\begin{itemize}
		\item {\bf Premise 1:} Every physical system is described by a functional $S$, whose evolution is such that $\Delta S=0$.
		\item {\bf Premise 2:} The variation of a functional $S$, that is $\delta S$, is one and only one.
		\item {\bf Premise 3:} Humans (and the entire universe) are physical systems.
		\item {\bf Conclusion 1:} The evolution of humans (and the entire universe) is one and only one.
		\item {\bf Conclusion 2:} Free will is impossible.
	\end{itemize}
\section{Paradox in quantum mechanics}

The basis of quantum mechanics appears to be the only hope to support free will, but:

\begin{itemize}
	\item How could be interpreted a measurement by an observer that observes the measurement made by another observer?
	\item An {\bf internal observer} should not affect  the system after a measurement, since an {\bf external observer} can consider the system and the internal observer as a system.
	\item Every system is described by very specific equations. Such a system might be measured producing (according to quantum mechanics) uncertainty and altering its behavior, but what about the entire universe? is it being measured? if  not, how can be uncertain its evolution?
	\item Let $\ket{\Psi_i}$ be the state of a system in relation with an {\bf internal observer} and let $\ket{\Psi_e}$ be the state of the system in relation with an {\bf external observer}. Suppose
	\begin{equation}
		\ket{\Psi_i}=\sum a_n \ket{\phi_n}
	\end{equation}
	and 
	\begin{equation}
		\ket{\Psi_e}=\ket{\Psi_p}+\sum a_n \ket{\phi_n}
	\end{equation}
	where $\ket{\Psi_p}$ represents the internal observer (which is another system from the point of view of the external observer). In the internal observer a measurement of a physical quantity is taken giving the result $a_n$, according to the postulates, the state of the system immediately after the measurement is the normalized projection 
	\begin{equation}
		\Psi\to\frac{P_n\ket{\Psi}}{\bra{\Psi}P_n\ket{\Psi}}
	\end{equation}
	A process also known as Wave Funciton Collapse. Thus, we have that the wave function in relation with the internal observer has collapsed after the measurement. But, since the external observer has not decided to measure, its wave function preserves its form, and there is still posibilities for different results in the measurement, which contradicts the internal observer. It should be taken into account the interaction between the internal observer and the system, but it can not be caculated as long as the internal observer has free will.

		 	\item The wave function or state of the system that describes the universe is ruled by physical laws, hence its evolution is predetermined, and the entire universe can not be measured by an external observer, as a result there is not uncertainty or wave function collapse whatsoever.
	
\end{itemize}
Quantum mechanics is giving a privilege charachter to humans. It is implicitly stated in the Wave function collapse. In addition, it appears that quantum mechanics does not describe particles per se, but different objects that behaves as wave functions, hence it is futile to force the theory to provide information out of its theoretical machinery. Let us consider this analogy: If it is seen a cylinder from a position, it appears to be a circle, but if it is seen from another position, it appears to be a rectangle, but it is after all something more, a cylinder. The electron is observed as a wave, but also as a particle, probably there is something beyond to describe its nature properly.

\section{On Justice and Ethics.}
It has been already stated that physics (the most fundamental science) asserts that every phenomena is entangled, having influence and consequences surround, in other words, causality is a fundamental principle whose primary implication is determinism. Under this assumption, given the entanglement of the entities in the universe, we may ask: are our actions just? or unjust? because we can not really have any control, it is already established by initial conditions. In this sense, a criminal can not be found culpable since the sorrounding conditions obliged the system to evolve in that direction. If the universe do not create the conditions, such a criminal damage would be imposible.

In the contex of ethics, how can we take responsibility of our actions when we are unable to make decisions? it is considering to someone morally responsible of an act that is beyond his power to evade.  Justice and ethics are irrelevant concepts through which nothing can be done.

\section{Conclusion}
Every system is described by a set of variables submitted to the principle of least action, which asserts that the system is described by a extreme value of a functional. Every functional has a unique solution, thus every system has but one way to evolve, in other words, every theory in physics lead to a deterministic description. The universe is a system submitted to the laws of physics, in consequence, the universe has but one way to evolve, in a metaphorical terms, the destiny of the universe has been sealed. 

It seems that the reason of determinism, from a metaphysical point of view, is the presence of entanglement and causality, ruling the configuration of the system. Yet, remain many questions, as for instance, why and how do we think?

On the other hand, it might be argued that the uncertainty principle allows the existence of choice due to the non-existence of determinism, also opening the possibility of choice and even casualty, but that it is primarily due to the probabilistic character of the theory (or the probabilistic approach of quantum mechanics which is being abandoned by physicist without knowing it) something that already adverted by Einstein. Such a principle seems to lead to paradoxes. Suppose a system and an external observer, according to the uncertainty principle the measurements made by the observer are inaccurate, but now imagine a system formed by a system and an observer, how are the measurements of the {\bf internal observer} in relation with the external observer? What about the state function of the entire universe? there is not external observer in relation with the universe, therefore its evolution cannot be altered.

Finally, the physical systems ruled by casualty and luck are inscrutable. It does not make any sense to research them.\\

``The fact that the path of an electron can not be determined does not imply that such a path does not exist".


%

\ifCLASSOPTIONcaptionsoff
  \newpage
\fi

\end{document}